\documentclass[8pt,aps,twocolumn]{revtex4}
\usepackage{amssymb}
\usepackage{latexsym}

\usepackage{epsfig}

\usepackage{dcolumn}
\usepackage{amsmath}
\usepackage{amsfonts}
\usepackage{bm}

\newcommand{\be}{\begin{equation}}
\newcommand{\ee}{\end{equation}}
\newcommand{\bea}{\begin{eqnarray}}
\newcommand{\eea}{\end{eqnarray}}

\newcommand{\p}{\partial}
\newcommand{\s}{\sigma}

\newcommand{\la}{\langle}
\newcommand{\ra}{\rangle}
\newcommand{\rd}{\mbox{d}}
\newcommand{\ri}{\mbox{i}}
\newcommand{\re}{\mbox{e}}

\begin{document}
\title{Riding a wild horse: Majorana fermions interacting with solitons of fast bosonic fields.}

\author{   A.  M. Tsvelik}
\affiliation
{ Department of  Condensed Matter Physics and Materials Science, Brookhaven National Laboratory, Upton, NY 11973-5000, USA}

\begin{abstract}I consider a class of one-dimensional models where Majorana fermions interact with bosonic fields. Contrary to a more familiar situation where bosonic degrees of freedom are phonons and as such form a slow subsystem, I consider fast bosons. Such situation exists when the bosonic modes appear as collective excitations of interacting electrons as, for instance, in superconductors or carbon nanotubes. It is shown that an entire new class of excitations emerge, namely bound states of solitons and Majorana fermions. The latter bound states are not topological and their existence and number depend on the interactions and the soliton's velocity. Intriguingly the number of bound states increases with the  soliton's velocity. 
\end{abstract}


\maketitle

\section{Introduction}

 Models describing Majorana fermions have received a lot of attention, mostly in the context of quantum computation. In the process of the discussion several models with interesting common features have emerged. In all these models Majorana fermions interact with solitons of bosonic fields representing collective degrees of freedom of electronic systems (usually  phase fields of superconducting order parameters). Although models of interacting solitons and fermions have been considered before and there is an extensive literature on the subject, there are some new features which merit a discussion. The most important new feature shared by the models in question is that Majorana fermions are slow in comparison to the bosonic modes. This is opposite to a more familiar situation of the  Peierls-Froelich model of polyacetylene  where a slow   optical phonons interact with fast  fermions. 

 As I will  demonstrate, in this situation new branches of fermion-soliton bound states  emerge some of which exist only in a finite region of momentum space. 

\section{Models}

\subsection{p-wave Josephson junction}

  This  model was formulated by Grosfeld and Stern \cite{stern} who considered  a long insulating one-dimensional Josephson junction between two p-wave superconductors.  As it is well known,  p-wave superconductors have zero energy Majorana modes as boundary states. When the boundary is extended, as in the case of a long junction, these modes propagate along the edge with  velocity $v_{\Delta} \sim \Delta$, the superconducting gap in the bulk. A conventional long Josephson junction is described by the sine-Gordon model; the p-wave one acquires an additional term in the Hamiltonian corresponding to Majorana fermions. The resulting model of the junction has the following  Lagrangian density:
  \bea
 && {\cal L} = \label{ssg}\\
 && \frac{\tilde c}{2\beta^2}\Big[ {\tilde c}^{-2}(\p_{\tau}\Phi)^2 + (\p_x\Phi)^2 + \frac{4}{\lambda_J^2}\sin^2(\Phi/2)\Big] + \nonumber\\
 && \frac{\ri}{2}\Big(r\p_{\tau}r - vr\p_x r + l\p_{\tau} l + vl\p_x l\Big) + \ri m rl\cos(\Phi/2), \nonumber
  \eea
  where $r,l$ are right- and left-moving Majorana fermion fields propagating along different sides of the junction and $\Phi$-field represents the phase difference between the two superconductors. The parameters of the model are
  \bea
  \tilde c = c\sqrt{\frac{d}{d +2\lambda_L}}, ~~ \beta^2/8\pi = \frac{2 e^2}{\hbar c}\sqrt{\frac{d(d + 2\lambda)}{h_z^2}},
  \eea
  where $d$ is the thickness of the barrier, $\lambda_L$ is the London penetration depth, $h_z$ is the hight of the junction and $\lambda_J$ and $m$ are determined by characteristics of the junction. The magnetic field changes the argument of the sinus to 
  \bea
  \Phi/2 \rightarrow \Phi/2 - eBd x/c\hbar.
  \eea
  Model (\ref{ssg}) is  a generalization of the Super sine-Gordon (SSG) model; for the latter case 
  \bea
  m = v/\lambda_J, ~~ \tilde c =v.\label{Lorentz}
  \eea
  so that the model is Lorentz invariant. The SSG model is integrable \cite{tsv} and its excitations and S-matrix are known \cite{ahn}. In particular, it is known that solitons of SSG model obey non-Abelian statistics.
  
  Since all candidates for p-wave superconductivity have small critical temperatures corresponding to small  $v$, the ratio $v/\tilde c$ is also likely to be very small (however, it may be increased somewhat by putting the junction on top of a dielectric with large $\epsilon$). 
  
 \subsection{Combination of spin-orbit interaction and superconductivity}
 
 Although this model is not qualitatively different from the previous one, there are certain features which make its material realization easier. Namely, the chiral Majorana modes  emerge here not from a $p$-wave superconductor, which remains a rather exotic object, but by other means.  Namely, Majorana fermions may emerge in a semiconductor with a strong spin-orbit interaction subject to external magnetic field brought into contact with an $s$-wave superconductor (see, for example \cite{oppen},\cite{das}). Following \cite{oppen} I write down the Hamiltonian of two-dimensional film of such material:
 \bea
 && H = \int \Psi^+(x,y){\cal H}\Psi(x,y)\rd x\rd y, \\
 && \Psi^+ = (\psi^+_{\uparrow},\psi^+_{\downarrow},\psi_{\downarrow}, -\psi_{\uparrow}),\nonumber\\
 && {\cal H} = [{\bf p}^2/2m - \mu]\tau_z + u(p_y\s^z - p_x\s^y)\tau_z + \nonumber\\
 && B(y)\s^x + \Delta(y)\tau_x,\nonumber
 \eea
 where $\tau_a$ act in particle-hole and $\s^a$ in spin space respectively. As was shown in \cite{oppen}, when function $V(y) = B(y) + \Delta(y)$ changes sign Majorana zero modes emerge. The corresponding operators of right- and left moving modes are made of combinations 
 \bea
 && r =\frac{1}{2}(\psi_{\uparrow} -\ri\psi_{\downarrow} + \psi^+_{\uparrow} + \ri\psi^+_{\downarrow}), \\
 && l = \frac{1}{2}(\psi_{\uparrow} +\ri\psi_{\downarrow} + \psi^+_{\uparrow} - \ri\psi^+_{\downarrow}),
  \eea
These modes are spatially separated: the mode $r$ emerges at the edge with $\rd V/\rd y >0$ and $l$-mode at the edge $\rd V/\rd y <0$. These edges are boundaries between the topological insulator and the superconducting state. Being projected onto these modes the Hamiltonian density becomes
\bea
H_0 =\frac{u}{2} \int \rd x (-r\p_x r + l\p_x l).
\eea
Since the spin-orbit interaction is typically much greater in magnitude than the $p$-wave order parameter, one can increase the ratio $u/\tilde c$. In \cite{oppen} the authors cite 
$u  = 7.6\times 10^6 cm/sec$ for InAr. With $\lambda_L \sim 10^4$A and $d \sim 10$A one can get $u/\tilde c \sim 10^{-2}$. 

 Now following \cite{oppen2} consider a situation when a narrow superconducting region is sandwiched between two topological insulators. Then instead of one bosonic mode as in (\ref{ssg}) we will have more.  Namely, if the superconducting region between two topological insulators is sufficiently narrow, we have the following action \cite{oppen2}:
 
 \bea
 && - J\cos(\phi_a - \phi_b) + \label{oppen}\\
 && \ri rl\Big[t_1\cos\Big(\frac{\phi_a - \phi_b}{2}\Big) +  t_2 \cos\Big(\frac{\phi_a + \phi_b}{2}- \phi_m\Big)\Big]\nonumber
 \eea
 where $\phi_{a,b}$ are superconducting phases on the left and right from the superconducting strip and $\phi_m$ is a phase on the strip. Therefore there are two independent bosonic modes.
 
 \subsection{Half filled carbon nanotube}
  
  In \cite{ners} the author and Nersesyan derived an effective field theory for armchair carbon nanotubes using the bosonization approach with a partial refermionization. At half filling the model is similar to (\ref{ssg}), but the number of Majorana fermions is not one, but 6 with different mass parameters $m_a$ such that 
  \bea
  && m_{-2} = m_{-1} =m_c, ~~ m_1=m_2 =m_3 = m_t, \nonumber\\
  &&  2m_c + m_t +m_0 =0.
  \eea
   The role of the bosonic field $\Phi$ is played by the total charge field $\Phi_c$. The smallness of parameter $\beta$ and a large value of $\tilde c/v$ originate from the unscreened Coulomb interaction.  The Lagrangian density is 
  \bea
 && {\cal L} = \frac{1}{2v}(\p_t\Phi_c)^2 - \frac{v}{2K^2}(\p_x\Phi_c)^2 + \nonumber\\
 && \frac{\ri}{2}\sum_{a=-2}^3\bar\chi_a\Big(\gamma_0\p_t + v\gamma_1\p_x\Big)\chi_a - {\cal V},\label{carbon}\\
 && {\cal V} = \\
 && \ri\cos[\sqrt{4\pi}\Phi_c]\Big[m_1\sum_{a=-2}^{-1}\bar\chi_a\chi_a + m_2\sum_{a=1}^3\bar\chi_a\chi_a + m_3\bar\chi_0\chi_0\Big],\nonumber
 \eea
 where $\gamma_0,\gamma_1$ are Dirac gamma matrices, $K << 1$ is the Luttinger parameter  determined by the long range Coulomb interaction, $\Phi_c$ is the total charge field and $\chi_a$ are Majorana fermions made of chiral components of $\Phi_f$  ($a =-2,-1$) and $\Phi_s,\Phi_{sf}$ fields ($a=0,1,2,3$) \cite{ners}. The ${\cal V}$-term represents the leading interaction generated by the Umklapp processes; the interactions between the Majorana fermions are small in comparison. The symmetry of (\ref{carbon}) is U(1)$\times$U(1)$\times$SU(2)$\times$Z$_2$. The Majorana modes with $a=1,2,3$ realize the S=1 representation of the SU(2) group. 
 
  For $K << 1$ $\Phi_c$ is essentially a classical field and its dynamics is determined by the equation
  \bea
 &&  - v^{-1}\p_t^2\Phi_c(t,x) + vK^{-2}\p_x^2\Phi_c(t,x) + \label{sineG}\\
 && \sqrt{4\pi}\sin[\sqrt{4\pi}\Phi_c(t,x)]\sum_a m_a\la \chi_a(t,x)\bar\chi_a(t,x)\ra =0, \nonumber
  \eea
 At $K << 1$ one can neglect coordinate dependence of the mass term and treat it as a constant for the purposes of calculation of the fermion average in  (\ref{sineG}). The result is 
 \bea
&&  - v^{-2}\p_t^2\Phi(t,x) + K^{-2}\p_x^2\Phi(t,x) + M^2\sin[\Phi(t,x)] =0,\nonumber\\
&& M^2 \approx \sum_a \frac{m_a^2}{ v^2}\ln(\Lambda/|m_a|),
 \eea
 where $\Phi = \sqrt{16\pi}\Phi_c$ and  $M^2$ is calculated with the logarithmic accuracy.  This description is valid for excitations moving with velocities $< v$.  So we see that the solitons of $\Phi_c$ are solutions of the sine-Gordon equation.  As far as the fermionic excitations are concerned, they are determined by the same equations as for model (\ref{ssg}).  In the notations of (\ref{ssg}) we have  
 \be
 1/\lambda_J = KM.
 \ee
 As is shown the subsequent Section, the fermions have bound states with the solitons such that   the number of finite energy bound states in each channel  is 
 \bea
 N_{0,a} = m_a\lambda_J/v =  \frac{m_a}{K\Big[\sum_b m_b^2 \ln(\Lambda/|m_b|)\Big]^{1/2}}.
 \eea
 It is clear that for the Lorentz invariant case $K=1$ there are no finite energy bound states. More than that, they appear only if the Coulomb interaction is quite strong.

  \section{Semiclassical analysis}
  
  In all models described above the bosonic action is of the sine-Gordon type. The sine-Gordon subsystem has two types of excitations: kinks and breathers. Kinks strongly interact with the Majorana fermions since the latter ones create bound states with kinks. There are two types of  bound states: one type is the Majorana zero modes which modify the kink's quantum numbers and the other are massive ones. Below I do the analysis for model (\ref{ssg}), generalizations for models (\ref{oppen},\ref{carbon}) are straightforward. In particular, in model (\ref{carbon}) $\tilde c = v/K$.
  
  It is convenient to introduce new fermionic fields:
  \be
  \chi_1 = (r+l)/\sqrt 2, ~~ \chi_2 = (r-l)/\sqrt 2
\ee
I consider a static kink configuration first. Then for a kink centered around $x_0$ the fermion operators can be represented by the mode expansion (\ref{modes})\cite{witten}. The eigenfunctions satisfy 
\bea
&& E\psi_1 = \ri(v\p_x - W)\psi_2, ~~ E\psi_2 = \ri(v\p_x + W)\psi_1, \nonumber\\
&& E^2\psi_1 = 
 (- v^2\p^2_x + W^2 - v\p_x W)\psi_1\label{dirac}
\eea
where $W = m\cos(\Phi/2)$. 

\bea
 && \Big(
 \begin{array}{c}
 \chi_1(x,t)\\
 \chi_2(x,t)
 \end{array}
 \Big) = \gamma_0\Big(
 \begin{array}{c}
 \psi_{0}(x-x_0)\\
 0
 \end{array}
 \Big) +  \label{modes}\\
 && \sum_{E_n >0} \Big\{\Big[\re^{-\ri E_nt}\hat\gamma_n(x_0) +\re^{\ri E_nt}\hat\gamma_n^+(x_0)\Big] \Big(
 \begin{array}{c}
 \psi_{1,n}(x-x_0)\\
 0
 \end{array}
 \Big) + \nonumber\\
 && \frac{\ri}{ E_n}\Big[\re^{-\ri E_nt}\hat\gamma_n(x_0) -\re^{\ri E_nt}\hat\gamma_n^+(x_0)\Big] \times\nonumber\\
 && \Big(
 \begin{array}{c}
 0\\
 (v\p_x +W)\psi_{1,n}(x-x_0)
 \end{array}
 \Big) \Big\}\nonumber
 \eea
A general single-soliton solution is 
\bea
\Phi = 4\tan^{-1}\Big[\exp\Big(\frac{x - x_0 - ut}{\lambda_J\sqrt{1 - (u/\tilde c)^2}}\Big)\Big].
\eea
In the static case $u=0$ this gives rise to the following potential:
\bea
W = m\tanh(x/\lambda_J).
\eea
Substituting it in (\ref{dirac}) and using the results from \cite{landau}, I obtain the following  eigenvalues for  the bound states:
\bea
&& E^2_n = \Big(\frac{2m}{N_0}\Big)^2 n(2N_0 - n), \nonumber\\
&& n=0,1,...  < N_0 = m\lambda_J/v. \label{energy}
\eea
and the eigenfunctions are
\bea
&& \psi_1 =  (\cosh\xi)^{(-N_0 +n)}\times\\
&& F\Big(-n,2N_0 - n+1,N_0 -n +1; \frac{1}{\re^{2\xi} +1}\Big), \nonumber\\
&&  \xi = x/\lambda_J.\nonumber
\eea
Notice that in the Lorentz invariant case (\ref{Lorentz}) $N_0 =1$ and the only bound state is the topological one $n=0$.

Now let us consider the case of moving soliton $u \neq 0$.  Let us introduce new coordinates:
 \bea
 && x' = \gamma (x-ut), ~~ t' = \gamma(t -xu/v^2), \nonumber\\
 && \gamma= [1-(u/v)^2]^{-1/2}.
 \eea
 This Lorentz transformation leaves the fermionic action invariant and puts us in the reference frame of the moving soliton. The mass term in (\ref{dirac})  becomes
 \bea
 W\Big(\frac{x'}{\lambda'}\Big), ~~ \lambda' = \lambda_J\Big[\frac{1 - (u/\tilde c)^2}{1- (u/v)^2}\Big]^{1/2}.
 \eea
 Notice that in the Lorentz invariant case $\tilde c =v$ the scale does not change. However, if $v < \tilde c$, the soliton size increases and, as a consequence, the number of bound states $N$ also increases:
 \bea
 N = N_0\Big[\frac{1 - (u/\tilde c)^2}{1- (u/v)^2}\Big]^{1/2}.
 \eea
 This is a somewhat unexpected result. The energy in this reference frame is given by (\ref{energy}) with $N_0$ replaced by $N$. 

In the laboratory reference frame the energy and momentum of the fermionic part of the bound state  are (I set $\tilde c \rightarrow \infty$):
\bea
&& E_n(u) = \frac{2m}{N_0}\sqrt{n\Big[\frac{2N_0}{\sqrt{1- (u/v)^2}} -n\Big]}, \label{masses}\\
&& P_n(u) = uE_n(u)/v^2, ~~n =0,...N_0/\sqrt{1- (u/v)^2}-1, \nonumber
\eea
 To obtain the total energy and momentum of the kink and the bound state one has to add the energy and momentum of the kink:
 \bea
 E_k(u) = \frac{M_k}{\sqrt{1 - (u/\tilde c)^2}}, ~~ P_k(u) = \frac{M_k u}{{\tilde c}^2\sqrt{1 - (u/\tilde c)^2}}. \label{soldisp}
 \eea
 
 Consider $n=0$ mode first. For zero velocity this mode always exists, but for finite velocities its existence is restricted by the condition $u <v$. Assuming that $\tilde c/v$ is so large that the kink's momentum is much smaller than the fermionic part and its dispersion is slow, we can set $E_k = M_k, P_k =0$. Then from (\ref{soldisp}) we extract its dispersion:
 \bea
 && E(p) = \sqrt{M_k^2 + (\tilde c p)^2}, \\
 && |p| < \frac{M_k}{\tilde c}[(\tilde c/v)^2 -1]^{-1/2} \approx vM_k/{\tilde c}^2.\nonumber
 \eea
 Thus the zero energy bound state always exists, though in a limited region of the Brillouin zone. 
 
 Now consider the finite energy bound states. They are not topological and their existence is conditional. Somewhat unexpectedly (\ref{masses}) shows  that the conditions for their existence improve when the kink's velocity increases towards $v$. There are solutions with $N_0 [1- (u/v)^2]^{-1/2} > n > N_0$ existing only for finite velocities (momenta). It is   illustrated on Fig. 1.

\begin{figure}
\begin{center}
\epsfxsize=0.45\textwidth
\epsfbox{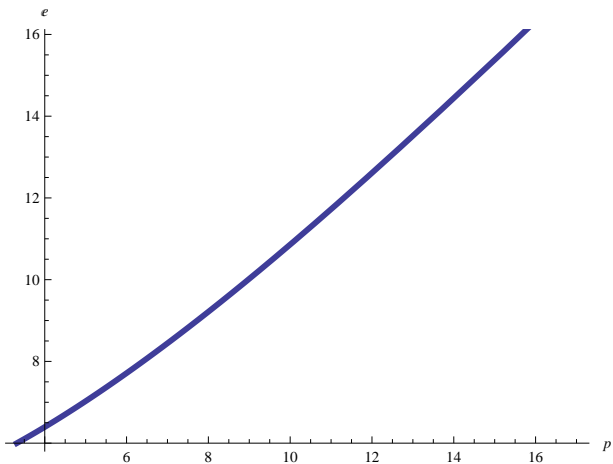}
\end{center}
\caption{ The spectrum $e =N_0[E(p)- M_k]/2m$ (\ref{masses}) for $N_0 =5$ and $n=6$. The spectrum exists only for momenta larger than critical as explained in the text. The fermion velocity is set $v=1$.} 
\label{rpa}
\end{figure}

To derive the dispersion of the bound states we have to take into account the fact that their energies and momenta are sums of (\ref{masses}) and (\ref{soldisp}). If we assume that  
\be
2m/N_0 >> M_k(v/\tilde c)^2,
\ee
then the inertia of  the kinks is very small and their contribution  to the total momentum can be neglected in comparison with the momentum of the bound state (\ref{masses}). As a result one   gets the picture of the dispersion depicted on Fig. 2. 

 \begin{figure}
\begin{center}
\epsfxsize=0.3\textwidth
\epsfbox{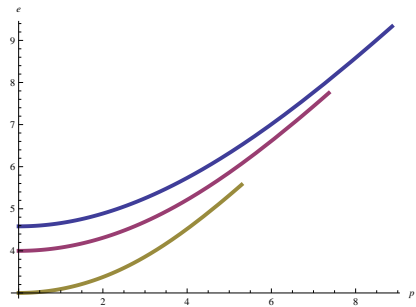}
\end{center}
\caption{The spectrum $e = N_0[E_n(p) - M_k]/2m$ (\ref{masses}) for $N_0 =5$ and $n=1,2,3$. The fermion velocity is set $v=1$.}
\end{figure}

 \section{Quantum numbers and correlation functions}
 
 As we see from (\ref{modes}) the operators of fermion-kink zero modes $\gamma_0(x_0)$  compose a Clifford algebra. For models with several species of Majorana fermions, such as (\ref{carbon}), $\gamma_0^a$ create a spinor representation of the corresponding group (for (\ref{carbon}) the group is O(6) $\sim$ SU(4) ) and the bound states of solitons and Majorana fermions transform according to this spinor representation. For the case of half filled carbon nanotube these excitations carry the same quantum numbers as the original fermions and therefore are  quasiparticles. The situation with $n \neq 0$ bound states is quite different. They transform according to the vector representation of  the corresponding group and therefore can be created only by pairs of fermionic operators. 
 
 In order to get a better grip of the picture, let us consider the model of carbon nanotube. The symmetry group of model (\ref{carbon}) is U(1)$\times$U(1)$\times$O(3)$\times$Z$_2$. 
 As an example of a local field  having nonzero matrix elements between the vacuum and the aforementioned bound states we have
\bea
 \re^{\ri\sqrt{4\pi}\varphi_{c}}\cos[\sqrt{4\pi}\varphi_f] \sim R^+_{1\uparrow}R^+_{1\downarrow} + R^+_{2\uparrow}R^+_{2\downarrow} ,
 \eea
 where $\varphi_{c,f}$ are right-moving components of the corresponding bosonic fields, $c$ labels the total charge,  $f$ labels the relative one and $1,2$ label positions of two Dirac points in the Brillouin zone of  carbon nanotube. The first exponent creates two right-moving solitons in the charge sector. One soliton creates a bound state with a Majorana fermion from $f$-sector (asymmetric charge) and the other soliton remains unbounded.  From this example one can see that the bound states can be observed only as parts of  continua. In the above example the continuum consists of a "naked" soliton and a soliton-fermion bound state. Existence of "naked" solitons, i.e. ones which do not carry any fermionic modes is guaranteed by the fact that $\tilde c > v$ and so their is plenty of room in momentum space for solitons which velocity exceeds the one of the fermions and those, as we know, do not create bound states. 
 
 \section{Conclusions}
 
  This paper demonstrates that in field theories without Lorentz invariance (quite a common thing in condensed matter physics) one has to expect appearance of new types of bound states, some of them existing only in a limited region of momentum space.  
 
 I am  grateful to Alexander Nersesyan and Robert Konik for interesting discussions. 
AMT was  supported  by US DOE under contract number DE-AC02 -98 CH 10886.


\begin{thebibliography}{99}
\bibitem{stern} E. Grosfeld and A. Stern, Proc. Natl. Acad. Sci. USA, 2011 Jul 19; {\bf 108}(29); 11810-4. 
\bibitem{tsv} A. Tsvelik,  Sov. J. Nucl. Phys. ( Yad. Fis. ) 47, 172 (1988).
\bibitem{ahn} C. Ahn, Nucl. Phys. B{\bf 354}, 57 (1991).
\bibitem{oppen} Y. Oreg, G. Refael and F. von Oppen, Phys. Rev. Lett. {\bf 105}, 177002 (2010). 
\bibitem{das} R. M. Lutchyn, T. D. Stanescu, and S. Das Sarma, Phys. Rev. Lett. {\bf 106}, 127001 (2011). 
\bibitem{oppen2} L. Jiang, D. Pekker, J. Alicea, G. Refael, Y. Oreg, and F. von Oppen, arXiv: 1107.4102.
\bibitem{ners} A. A. Nersesyan and A. M. Tsvelik, Phys. Rev. B{\bf 68}, 235419 (2003).
\bibitem{witten} E. Witten, Nucl. Phys. B{\bf 142}, 285 (1978).
\bibitem{landau} L. D. Landau and E. M. Lifshitz, "Quantum Mechanics (Non-relativistic theory)", Pergamon Press, 1977, pp. 73,74.


\end{thebibliography}
\end{document}